\documentclass[12pt]{article}
\usepackage{latexsym}
\usepackage{amssymb}
\usepackage{epsf}
\usepackage{amsmath}
\usepackage{cite}

\numberwithin{equation}{section}

\usepackage{amsmath,amssymb,epsfig,subfigure}
\usepackage[hypertex]{hyperref}
\usepackage{graphicx}

\begin{document}
\baselineskip=18pt
\pagestyle{plain}


%
%


\thispagestyle{empty}

\vspace*{-2cm}
\begin{flushright}
LA-UR-08-07575
\end{flushright}

\vspace*{1.7cm}
\begin{center}

{\bf \Large Supersymmetry breaking and gauge mediation in models with a generic superpotential} 
\vspace{1.3cm}


 \vspace*{1.5cm}
 { Ryuichiro Kitano$^{\dagger}$ and  Yutaka Ookouchi$^*$ }\\
 \vspace*{0.8cm}

$^\dagger$ {\it Theoretical Division T-2, Los Alamos National Laboratory, NM 87545, USA}\\
 $^*$
{\it Perimeter Institute for Theoretical Physics, ON N2L2Y5, Canada}\\
 
 \vspace*{0.8cm}
\end{center}
\vspace*{2.0cm}

\noindent
We present a general scheme for finding or creating a metastable vacuum
in supersymmetric theories.
%
%
%
By using the formalism, we show that there is a parameter region where a
metastable vacuum exists in the Wess-Zumino model coupled to messenger
fields. This model serves as a perturbative renormalizable model of
direct gauge mediation.


\newpage
\setcounter{page}{1} 



\setcounter{footnote}{0}

\section{Introduction}

If the theory to describe our universe possesses supersymmetry, the
vacuum of the theory should be somewhat unusual. Supersymmetry, which is
a part of the space-time symmetry, must be spontaneously broken in the
vacuum whereas the rest of the space-time symmetries, the Poincare
symmetry, are kept unbroken.
There have been many attempts to realize this hypothesis in
four-dimensional field theories. For example, the ideas of dynamical
supersymmetry breaking \cite{Witten,ADSI,ADSII} and gauge mediation \cite{DNS,GaugeMed2,GaugeMed3,GaugeMed4,GaugeMed5,GaugeMed6} (see also \cite{GRrev,Luty} for review and relevant references) have attracted much attention because of their beautiful
concepts.
However, a rather contrived set-up was necessary because generic
supersymmetric theories have a stable supersymmetric vacuum.



The recent discovery of metastable vacua in SQCD \cite{ISS} opened up
new avenue for a realistic model building because of its simplicity.
The idea of living in a metastable vacuum allows us to consider
vector-like models (for earlier works see \cite{ITIY}) whose
non-perturbative effects have been well understood \cite{IS}.
String-theory dualities can also provide useful tools to analyze such
theories and their vacuum structures.
%
%
Various works on this subject have been done (see e.g. \cite{ISII}),
including models with extended supersymmetry \cite{OOP,MOOP,Harv}
and geometrical realization of metastable vacua in string theories
\cite{IIA1,IIA2,IIA3,IIB1,IIB3,IIB4,IIB6,IIB7}. 
Also, the idea of metastable vacua drastically simplified a way of gauge
mediation and simple successful models have been proposed
\cite{KitanoI,KitanoII,DM,KOO,Terning,MurayamaNomuraI,MurayamaNomuraII,Ama,Extra,IK1,IK2,AJK,Haba}. Motivated by the fact that the low energy theories of many supersymmetry
breaking models are identical to the O'Raifeartaigh model, the vacuum of
generalized O'Raifeartaigh models have been studied in \cite{Ray,Shih,ShihII,FR,AKO,MS}.

The usual steps to build a gauge-mediation model are as follows. We
first construct/prepare a model of supersymmetry breaking. We introduce
messenger fields which carry quantum numbers of the standard-model gauge
group, and let them couple to the supersymmetry breaking model. (Or, if
possible we identify messenger fields in the supersymmetry breaking
model.) At this stage, we often encounter one of the following two
problems: the appearance of a new unstable direction towards a new
supersymmetric vacuum where the messenger fields acquire vacuum
expectation values, or vanishing gaugino masses. Therefore, in either
case, the final step is to deform the model so that the messenger fields
are stabilized and the gauginos obtain masses.
In short, we break supersymmetry, make it restored and try to break
again, or we break supersymmetry and deform the model while avoiding the
supersymmetry breaking vacuum to be destroyed. The first and the final
steps are quite simplified when we allow the supersymmetry breaking
vacuum to be metastable.
However, as one can notice from this discussion, it may be a detour to
start from a supersymmetry breaking model once we allow the
metastability. There can be a possibility that one starts from a generic
theory including the messenger fields and finds a (meta)stable minimum
of the potential.


In this note, we present a transparent scheme for finding or creating a
(meta)stable vacuum in general supersymmetric models.
%
We derive general conditions for having a supersymmetry breaking vacuum
by connecting different models by a coordinate transformation, which is
an application of the method used in \cite{OOP}.
%
%
In particular, we find that there can be a metastable supersymmetry
breaking vacuum in models with the canonical K\"ahler potential and a
generic superpotential. For example, the Wess-Zumino model coupled to
the messenger fields possesses a metastable vacuum if coefficients of the
superpotential terms satisfy certain inequalities.


\section{Local equivalence to the Polonyi model}

In this section we develop a general method of creating a metastable
supersymmetry breaking vacuum. Let us start with the following
Lagrangian defined by
\begin{eqnarray}
{\cal L} = [K(z,\bar z)]_D - \left( [\epsilon \, z ]_F + {\rm h.c.}
			     \right),
\nonumber 
\end{eqnarray}
where $z$ is a chiral superfield and $\epsilon$ is a parameter.
This model (the Polonyi model \cite{Polonyi:1977pj}) has a stable SUSY breaking vacuum at
$z=0$ if
\begin{eqnarray}
 K_{z \bar z}|_{z=0} > 0,
\label{eq:cond1}
\end{eqnarray}
\begin{eqnarray}
 K_{z z \bar z}|_{z=0} = K_{z \bar z \bar z}|_{z=0} = 0,
\label{eq:cond2}
\end{eqnarray}
\begin{eqnarray}
 K_{z z \bar z \bar z} |_{z=0} < 0,
\label{eq:cond3}
\end{eqnarray}
\begin{eqnarray}
 K_{z z \bar z \bar z}^2 |_{z=0} - \left|  K_{z z z \bar z} |_{z=0}
				   \right|^2 > 0.
\label{eq:cond4}
\end{eqnarray}
We implicitly assumed that the function $K$ is differentiable at least
four times at $z=0$. The condition in (\ref{eq:cond1}) is a sign
convention of the kinetic term.
As long as the sign is correct, we can normalize the $z$ field so that
$K_{z \bar z}|_{z=0} = 1$.
Also, (\ref{eq:cond2}) can always be satisfied by an appropriate
shift of $z$.
With this convention, the conditions (\ref{eq:cond3}) and
(\ref{eq:cond4}) ensure the stability of the potential. 
%
For convenience, we define
\begin{eqnarray}
 \alpha \equiv K_{z z \bar z \bar z} |_{z=0},\ \ \ 
 \beta \equiv K_{z z z \bar z} |_{z=0},\ \ \ \nonumber
\end{eqnarray}
with the canonical normalization of the field, $K_{z \bar z}|_{z=0} =
1$.  The $\alpha$ parameter is a real number whereas $\beta$ is a
complex number.
The masses of the scalar components, $m_R$ and $m_I$, satisfy
\begin{eqnarray}
 m_R^2 + m_I^2 = - 2 |\epsilon|^2 \alpha,\ \ 
 m_R^2 m_I^2 = |\epsilon|^4 (\alpha^2 - |\beta|^2). \nonumber
\end{eqnarray}
The vacuum energy at $z=0$ is
\begin{eqnarray}
 V|_{z=0} = |\epsilon|^2. \nonumber
\end{eqnarray}

Now, we perform a field redefinition (or a coordinate transformation),
$z \to X = w^{-1} (z)$, where $w^{-1}(z)$ is the inverse of a
holomorphic function $w(X)$. The function $w(X)$ is required to be
analytic in the neighbor of $X=x_0$ where $w(x_0)=0$, and also the first
derivative should not vanish ($w_X|_{x_0} \neq 0$) in order for the
inverse function to be at least locally defined around the stable point
$z=0$. Since we are interested only in the local structure of the
potential, we do not need the transformation to be globally defined.
The superpotential in terms of $X$ is given by
\begin{eqnarray}
 W = \epsilon w(X).
\label{eq:superX}
\end{eqnarray}
This transformation generates various kinds of (metastable) SUSY
breaking model by simply choosing an arbitrary holomorphic function
$w(X)$. Here, for simplicity, we restrict ourselves to the case that
$w(X)$ is a function of a single chiral superfield, $X$. In a more general
case, the discussion can be reduced to the single-field case by
integrating out other degrees of freedom.
In appendix we present a general discussion.

The stable point at $z=0$ maps to $X=x_0$. By assumption, we can expand
the function $w(X)$ around $x_0$:
\begin{eqnarray}
 w(X) = \xi(x_0) \left( \Delta X + {\xi_2(x_0) \over 2} \Delta X^2 
+ {\xi_3(x_0) \over 6} \Delta X^3 + \cdots \right),
\end{eqnarray}
where $\xi(x_0) = w_X|_{x_0} \neq 0$, $\xi_2(x_0) = w_{XX}|_{x_0}/\xi(x_0)$ and $\xi_3(x_0) =
w_{XXX}|_{x_0}/\xi(x_0)$, and $\Delta X = X-x_0$.
We can take $\xi(x_0) = 1$ which corresponds to the change of the function:
$w(X) \to w(X)/\xi(x_0)$.
The higher order terms of $O(\Delta X^4)$ are irrelevant for the discussion of
the local stability.
The expansion of the K\"ahler potential in terms of the $X$ field around
$X = x_0$ has the following coefficients:
\begin{eqnarray}
 K_{X X \bar X}|_{x_0} = \xi_2(x_0),
\label{eq:eq2}
\end{eqnarray}
\begin{eqnarray}
 K_{X X \bar X \bar X}|_{x_0} = |\xi_2(x_0)|^2 + \alpha,
\label{eq:eq3}
\end{eqnarray}
\begin{eqnarray}
 K_{X X X \bar X}|_{x_0} = \xi_3(x_0)  + \beta  .
\label{eq:eq4}
\end{eqnarray}
Here we take the canonical normalization of the $X$ field, $K_{X \bar
X}|_{x_0} = 1$.  Higher order terms are again irrelevant for the
discussion.
The meaning of the equations is the following. With a given
superpotential and a K\"ahler potential for the $X$ field, there is a
(meta)stable minimum at $X=x_0$ if it is possible to find $\alpha$,
$\beta$, and $x_0$ which satisfy the above equations and the conditions
in (\ref{eq:cond3}) and (\ref{eq:cond4}), i.e.,
\begin{eqnarray}
-\alpha > |\beta|.
\label{eq:alphabeta}
\end{eqnarray}
For $\beta=0$, the field redefinition $X \to w(X)$ defined by the
coefficients, $\xi_{2,3}$, from (\ref{eq:eq2}) and (\ref{eq:eq4})
corresponds to the transformation to the K\"ahler normal
coordinate \eqref{KNC} in our normalization.

Since there are five real equations for five real variables
 ($\alpha$, Re$[\beta]$, Im$[\beta]$, Re$[x_0]$, Im$[x_0]$), there is
 generically a solution.
These equations are particularly useful when we know the K\"ahler
potential. In such a case, one can choose a point $x_0$, set $\beta =
0$, and then fix $\xi_2$, $\xi_3$ and $\alpha$ by using the above
equations.\footnote{Or, one can fix K\"ahler terms if the superpotential
is known.} If $\alpha$ is negative, we can obtain a superpotential which
creates a metastable vacuum at $X=x_0$.
Note that this is not a fine-tuning of the parameters. Around
the constructed solution, there is a region of the parameter space where
a (meta)stable vacuum exists.
%
%

The coordinate transformation $z \to w^{-1}(z)$ can modify the global
structure of the potential although the local stability is ensured in
this formulation. For example, the Wess-Zumino model obtained by this
transformation has SUSY vacua whereas the original Polonyi model does
not.
Conversely, if the transformation is globally defined, there is no SUSY
vacuum because two models are the same.
In the metastable case, the lifetime of the vacuum can be arbitrarily
long by taking $\epsilon$ to be small.

In general, this formulation can be used to find a metastable vacuum in
a given model by checking if the model is locally equivalent to the
Polonyi model, i.e., if a solution is constructable.
%
%
By following the derivation inversely, one can see that the above
condition is a necessary and sufficient condition for supersymmetry
breaking.
%
%
%
%
%
A necessary condition for having a metastable vacuum
is therefore an existence of a point $x_0$ where
\begin{eqnarray}
 K_{X X \bar X \bar X}|_{x_0} < 
\left|
 K_{X X \bar X}|_{x_0}
\right|^2.
\end{eqnarray}

An interesting observation is that there is a useful sufficient
condition for the metastability. If the K\"ahler potential satisfies
\begin{eqnarray}
 K_{X X \bar X \bar X} = 0
\label{eq:double}
\end{eqnarray}
everywhere (except singular points) in the $X$ space, one can create a
metastable vacuum anywhere by adding a suitable superpotential. $N=2$
supersymmetric theories are such examples \cite{OOP}.

\section{Gauge Mediation Models}

\subsection{General argument}

In this section we will construct models of direct gauge mediation where
the particle content is supersymmetry breaking fields, $X_i$, and the
messenger fields, $f_k,\tilde{f}_k$. We take the canonical
K\"ahler potential: $${
K=X^i{X}_i^{\dagger}+f^kf^{\dagger}_k+\tilde{f}^k\tilde{f}^{\dagger}_k.
}$$
Suppose the superpotential includes interaction terms,
\begin{eqnarray}
{\cal W}=M(X)^{kl}f_k\tilde{f}_l +\epsilon\, W(X),
\end{eqnarray}
where $M(X)$ and $W(X)$ are arbitrary holomorphic functions of
$X_i$. For successful gauge mediation, the messengers should acquire
masses at the vacuum. Thus we are interested in a parameter region where
all the eigenvalues of the mass matrix $M(X)$ are non-vanishing and much
larger than  that of the field $X_i$ and $\sqrt{ F_{X^i} }$. To achieve this, we take $\epsilon$ to
be small, $\epsilon \ll 1$. At the low energy, the massive messengers can be
integrated out and correction terms to the K\"ahler potential of the fields $X_i$ are generated:
$${
K=K_{cl}+\eta K_{1loop}+\eta^2 K_{2 loop}+\cdots
}$$
where $\eta$ is one loop factor. They yield non-zero curvature in the $X$ space, which allows us to use
the method shown in the previous section.

Suppose that there is a point $x_0$ where the sectional curvature is postive definite, (for a model with single $X$, $K_{X\bar{X}X\bar{X}}|_{x_0} <
| K_{XX\bar X}|_{x_0} |^2$). Then as we have seen, by taking the
superpotential $W(X)$ appropriately, we can create a metastable vacuum
at the point $x_0$. At the supersymmetry breaking minimum, $${ \langle
X^i \rangle =x^i_0+\theta^2 F_{X^{i}}, }$$ the messenger masses are
easily read off from the original Lagrangian,
\begin{eqnarray}
{\cal W}= \left[ M(x_0) +\theta^2 \, F_{X_i} \partial_i M(x_0)  \right]^{kl} f_k \tilde{f}_l+\cdots
\end{eqnarray}
To study the idea in more detail, let us focus on a
simple example.

\subsection{Simplest example}

We consider the case with a single $X$ field and a pair of $f$ and
$\tilde f$.  Take $M(X)$ be linear and $W(X)$ be a cubic function of $X$. The messenger is
${\bf 5}+\bar{{\bf 5}}$ in $SU(5)$.  $${ {\cal W}=\lambda X
f\tilde{f}+W_{\rm cubic}(X) .  }$$ This is the general renormalizable
theory with the singlet field $X$ and the messenger fields $f$ and
$\tilde f$.
The bare mass term $m f \tilde f$ can be eliminated by a shift of the
$X$ field. After integrating out the messenger fields, the K\"ahler potential of the $X$ is modified and the metric of it can be written as
$${
g\simeq e^{\log Z_X(X)}.
}$$
As in \cite{Sudano} by expanding the wavefunction renormalization in small expansion paramters $\eta\ll \eta \log {|X|\over \Lambda} \ll 1$, the effective K\"ahler
metric is given by
\begin{eqnarray}
 g(X) = g_0(\mu) \left( 1+ (\eta A^{(1)} +\eta^2 B^{(1)})\log {|X|^2\over \mu^2} +\eta^2 A^{(2)} \left( \log {|X|^2\over \mu^2}\right)^2
+\cdots  \right),
\label{eq:exampleK}
\end{eqnarray}
where the cut-off $\Lambda$ is renormalized by $g_{0}$. 
In this specific model, the one-loop factor $\eta$ defined above is $N_m |\lambda|^2/16\pi^2$ where $N_m$ is number of messengers, which is $5$ in our present model. To controll our metastable state, we have to keep two-loop correction terms. For computation of these correction terms, we take an advantage of thinking large $X$ region \cite{WittenHie,GiveonShih,Sudano}. Since we are interested in a parameter region where mass of $X$ is very small compared to the one of the messengers, which guarantee that vev of $X$ is much larger than SUSY breaking scale, $\langle X\rangle^2 \gg F$. In this case, we need only to know dominant log contributions. The other terms are suppressed by $F/\langle X \rangle^2$. These log contributions can be computed by essentially following the paper \cite{GR,AGLR,Sudano}. The $A^{(1)}$ and $A^{(2)}$ are computed respectively by discontinuity of the one-loop anomalous dimension and discontinuity of the derivative of it with respect to the RG time, $t=\log \mu$ at the threshold scale $\mu= x_0$. On the other hand, the coefficient $B^{(1)}$ is given by a discontinuity of the two-loop anomalous dimension. Using the known formulae for one-loop and two-loop anomalous dimension (see for example \cite{MartinVaughn}),
$${
\gamma^{\rm one}_X={N_m |\lambda|^2 \over 16\pi^2},\qquad \gamma_X^{\rm two}=-{N_m |\lambda|^4 \over (16\pi^2)^2},
}$$
where $\gamma_X\equiv -{1\over 2}\log Z_X$, we can compute the coefficients in the K\"ahler metric \eqref{eq:exampleK} explicitly, 
\begin{eqnarray}
\eta A^{(1)}&=&\Delta \gamma^{\rm one}_X=-\eta ,\nonumber \\
\eta^2 B^{(1)}&=&\Delta \gamma^{\rm two}_X=\eta^2 {1\over N_m}, \\
\eta^2 A^{(2)}&=&{1\over 2} (\Delta \gamma_X^{\rm one})^2+ {1\over 4} \Delta \left( {\partial \gamma^{\rm one}_X \over \partial t}\right) =-\eta^2 {1\over N_m},\nonumber
\end{eqnarray}
where we define $\Delta { O}\equiv O(t_{x_0}^{(-)})-{ O}({t_{x_0}^{(+)}})$. 
From this, we see that the sign of $A^{(2)}$ is negative, thus the condition $K_{X\bar{X}X\bar{X}}|_{x_0}<|K_{XX\bar{X}}|_{x_0}|^2$ is hold everywhere in the field space. Following the method shown in section $2$, we can create a metastable supersymmetry breaking vacuum anywhere we want in field space. 
The expansion coefficients around $X=x_0$ are
\begin{eqnarray}
 K_{X X \bar X} |_{x_0} = {\eta \over x_0}  A^{(1)} +{\cal O}(\eta^2) , \nonumber
\end{eqnarray}
\begin{eqnarray} 
 K_{X X \bar X \bar X} |_{x_0} = {2\eta^2 \over |x_0|^2}A^{(2)} +{\cal O}(\eta^3), \nonumber
\end{eqnarray}
\begin{eqnarray}
 K_{X X X \bar X} |_{x_0} = -{\eta \over x_0^2} A^{(1)}   +{\cal O}(\eta^2), \nonumber
\end{eqnarray}
where the $X$ field is canonically (re)normalized, $K_{X \bar X}|_{x_0}
= 1$.  A metastable vacuum at $X=x_0$ can be created by adding a
superpotential:
\begin{eqnarray}
 W_{\rm cubic} =  
c_1 (X-x_0)
+ {c_2 \over 2} (X-x_0)^2
+ {c_3 \over 6} (X-x_0)^3
,
\label{eq:exampleW}
\end{eqnarray}
with
\begin{eqnarray}
 {c_2 \over c_1} = K_{XX\bar{X}}|_{x_0},
\label{eq:sol1}
\end{eqnarray}
\begin{eqnarray}
 \left| {c_3 \over c_1} -K_{XXX\bar{X}}|_{x_0} \right|
< \left| K_{XX\bar{X}}|_{x_0} \right|^2\ -K_{X\bar{X}X\bar{X}}|_{x_0}.
\label{eq:crange}
\end{eqnarray}
The value of $c_1$ can be arbitrary. 

We can discuss the parameter region in terms of a
set of parameters whose mass dimensions are more transparent:
\begin{eqnarray}
 { W}_{\rm cubic} = m^2 \left(
 X + {X^2 \over 2 M} + {\kappa X^3 \over 6M^2}
\right).
\label{eq:supersuper}
\end{eqnarray}
%
%
The values of the dimensionful parameters $m$ and $M$ are arbitrary as
long as $m \ll M$.
For $\eta \ll 1$, the parameter $\kappa$ needs to be $O(\eta^{-1})$ and
the VEV $x_0$ is of $O(\eta)$ in the unit of $M$.
The condition in (\ref{eq:crange}) tells us that the $\kappa$
parameter should be within a narrow range: 
$\kappa = 1/(4 \eta) + 3/8+1/4 N_m \pm
O(\eta)$.\footnote{A naive estimation from (\ref{eq:crange}) gives
$\kappa = O(1/\eta) \pm O(1)$. However, there is a cancellation between
coefficients in the calculation from (\ref{eq:exampleW}) to
(\ref{eq:supersuper}). If we include an $X^4$ term, the $\kappa$
parameter can be arbitrary. However, the same degree of fine-tuning is
necessary between $\kappa$ and the coefficient of the $X^4$ term.} 
One needs a rather large $\eta$ to avoid the fine-tuning.



For a not very small $\eta$, i.e., for a not very large $\kappa$, the
hierarchical structure of the parameters in (\ref{eq:supersuper}) can
naturally be obtained, for example, from a theory with a small breaking
of $R$-symmetry. If we assign an $R$-charge $x$ to the $X$ field and
introduce a spurion for the $R$-breaking $\phi_R$ with the same charge
$x$, then the $R$-invariant superpotential has the following structure:
\begin{eqnarray}
 W_{\rm cubic} \sim \left(
\phi_R \over \Lambda
\right)^{(2 - 3 x) / x}
\left(
\phi_R^2 X + \phi_R X^2 + X^3
\right),
\end{eqnarray}
with $O(1)$ coefficients.\footnote{We assumed that there is no terms
with inverse powers of $\phi_R$.} The scale $\Lambda$ is a cut-off of the
theory. For $\phi_R \ll \Lambda$ and $x < 2/3$, the above superpotential
has the desired structure where $M \sim \phi_R$ and $m^2 \sim M^2
(\phi_R / \Lambda)^{(2-3x)/x}$.

\section*{Acknowledgements}

We would like to thank P. Argyres, F. Cachazo, H. Ooguri, K. Maruyoshi, J. Marsano for discussions. Research at Perimeter Institute for Theoretical Physics is supported in part by the Government of Canada through NSERC and by the Province of Ontario through MRI. YO would like to thank the University of Cincinnati and University of Pisa for kind hospitality.

\appendix

\setcounter{equation}{0}
\renewcommand{\theequation}{A.\arabic{equation}}
\appendix

\section{Model with multiple chiral superfields $X_i$}

In this section, we generalize the argument shown in section $2$ to models with multiple chiral superfields $X_i$ applying the method used in \cite{OOP}.  Suppose the K\"ahler potential is a generic function of chiral superfields $K(X_i)$ and pick a point $X^{i}_{0}$ in the $X$ space. In the K\"ahler normal coordinate \cite{Normal1,Normal2}, 
\begin{eqnarray}
\omega^i= \Delta X^i +{1\over 2} {\Gamma^i}_{jk}\big| \Delta^j \Delta X^k +{1\over 6}g^{m\bar{l}}\partial \Gamma_{\bar{l}ij}\big| \Delta X^i \Delta X^j \Delta X^k +\cdots , \label{KNC}
\end{eqnarray}
where $\Delta =X-X_0$, the inverse of the metric is 
$${
g^{i \bar{j}}=g^{i \bar{j}} \big| +{R^{i\bar{j}}}_{k\bar{l}}\big|\, \omega^k \bar{\omega}^{\bar{l}}+{\cal O}(\omega^3). 
}$$
Therefore as long as the sectional curvature is positive definite we can make a metastable state by adding the superpotential $W(X)=k_i \omega^i$.

It would be useful to find a {\it sufficient condition} for the positivity. In general, the curvature of the K\"ahler manifold is 
$${
{R^s}_{ik\bar{l}}=g^{s\bar{j}}g^{m\bar{n}}\, \bar{\partial}_{\bar{j}}g_{m\bar{l}} \, \bar{\partial}_{\bar{n}}g_{i\bar{k}}-g^{s\bar{j}}\, \bar{\partial}_{\bar{j}}\partial_{i}g_{k\bar{l}}.
}$$
Therefore when the second term at a point $X_0$ is vanishing, the potential becomes
$${
V=g^{k\bar{l}}k_k \bar{k}_{\bar{l}} + g^{k\bar{l}} (k_i\partial_i g_{s \bar{l}}\omega^{s} ) (\bar{k}_j\bar{\partial}_{\bar{j}} g_{k \bar{m}}\bar{\omega}^{\bar{m}} )+{\cal O}(\omega^3).
}$$
Thus, we obtain a sufficient condition for the stability
\begin{eqnarray}
\partial_i \bar{\partial}_{\bar{j}} g_{k\bar{l}}\big|_{X_0}=0 \quad \& \quad k_i \partial_{i} g_{j\bar{k}}\big|_{X_0} \neq 0  \qquad {\it sufficient \ condition}. \label{suf}
\end{eqnarray}
When the first equation holds everywhere (except singular points) in the $X$ space, then the metric can be written in terms of a holomorphic function. Equivalently the kinetic term can be written as
$${
{\rm Im}\int d^4 \theta\, {X^i}^{\dagger }{\partial {\cal F}(X) \over \partial X^i} .
}$$
This is the case of Seiberg-Witten theories \cite{SW}. As was shown in \cite{MOOP}, in this case the K\"ahler normal coordinate can be written in terms of the ${\cal F}(X)$,  
$${
\omega^i = X^i + m^{ij} {\partial {\cal F} \over \partial X_j} 
}$$
where $m^{ij}$ satisfies $1+{\bar{m}^{ij}} \partial_i \partial_j {\cal F}(X_0)=0$. The vacuum obtained in this way breaks all supersymmetry unless the theory have an extended supersymmetry \cite{MOOP}.

%
%

\end{document}